\documentclass[preprint,showpacs,aps]{revtex4}
\usepackage{graphicx}
\begin{document}
\def\be{\begin{equation}}
\def\ee{\end{equation}}
\def\bea{\begin{eqnarray}}
\def\eea{\end{eqnarray}}
\title{Geometrothermodynamics of black holes in two dimensions}
\author{Hernando Quevedo}
\email{quevedo@nucleares.unam.mx}
\affiliation{ 
Dipartimento di Fisica, Universit\`a di Roma "La Sapienza", Piazzale Aldo Moro 5, I-00185 Roma, Italy\\ 
ICRANet, Piazzale della Repubblica 10, I-65122 Pescara, Italy.}
\thanks{On sabbatical leave from Instituto de Ciencias Nucleares, Universidad Nacional Aut\'onoma de M\'exico}
\author{Alberto S\'anchez}
\email{asanchez@nucleares.unam.mx}
\affiliation{ Instituto de Ciencias Nucleares\\
Universidad Nacional Aut\'onoma de M\'exico  \\
AP 70543, M\'exico, DF 04510, Mexico}
\begin{abstract}

We study the properties of two-dimensional dilatonic
black holes from the viewpoint of geometrothermodynamics. 
We show that the thermodynamic curvature of the equilibrium
space vanishes only in the case of a flat spacetime,  
and it reproduces correctly the behavior of the thermodynamic interaction
and phase transition structure of the corresponding black hole configurations.

\end{abstract}

\maketitle

\section{Introduction}
\label{sec:int}
For the study of  the geometrothermodynamics (GTD) \cite{quev07} of black holes, the thermodynamic 
phase space ${\cal T}$ is assumed to be coordinatized by the set of independent coordinates 
$\{\Phi, E^a,I^a\}$, $a=1,...,n$, where $\Phi$ represents the thermodynamic potential, and $E^a$ and $I^a$ are the 
extensive and intensive thermodynamic variables, respectively. The positive integer $n$ indicates the number of
thermodynamic degrees of freedom of the black hole configuration. Moreover, the phase space is 
endowed with the Gibbs one-form 
$\Theta = d\Phi - \delta_{ab} I^a d E ^b$, $\delta_{ab} = {\rm diag}(1,...,1)$,
 and a particular metric
\be
G= (d\Phi - \delta_{ab} I^a d E^b)^2 +  (\delta_{ab}E^a I^b)(\eta_{cd} dE^c dI^d)
\ ,\qquad \eta_{ab}={\rm diag}(-1,1,...,1) \ ,
\label{gup}
\ee
which is invariant with respect to Legendre transformations $ \{\Phi, E^a,I^a\}\rightarrow \{\tilde \Phi, \tilde E ^a, \tilde I ^ a\}$,
with 
$ \Phi = \tilde \Phi - \delta_{ab} \tilde E ^a \tilde I ^b \ ,\ 
 E^a = - \tilde I ^ {a}, \  I^{a} = \tilde E ^ a $, \cite{arnold}. The equilibrium space ${\cal E} \subset {\cal T}$  is defined by the map $\varphi: 
  {\cal E} \rightarrow {\cal T}$ or, in local coordinates, $\varphi: \{E^a\}  \mapsto \{\Phi, E^a, I^a\}$, satisfying the condition 
$\varphi^* (\Theta) =0 $, i.e., on ${\cal E}$ it holds the first law of thermodynamics, $d\Phi = \delta_{ab} I^a d E ^b$, and the 
conditions of equilibrium $I^a = \delta^{ab}\partial \Phi/\partial E^b$ which relate the extensive variables $E^a$ with the intensive 
ones $I^a$. Then, the pullback $\varphi^*$ induces on ${\cal E}$, by means of $g=\varphi^*(G)$, the thermodynamic metric 
\be
g=\left(E^c\frac{\partial \Phi}{\partial E^c}\right)
\left(\eta_{ab}\delta^{bc}\frac{\partial^2\Phi}
{\partial E^c \partial E^d}
dE^a dE^d\right) \ .
\label{gdown}
\ee 
For the construction of this thermodynamic metric it is only necessary to know explicitly the thermodynamic potential in terms of the 
extensive variables $\Phi=\Phi(E^a)$. In black hole thermodynamics, the total mass $M$ is usually considered as the thermodynamic potential 
(canonical ensemble) in the energy representation 
and the fundamental equation $M=M(E^a)$ can be obtained from the area-entropy relationship $S= A/4$. 

In previous works we have shown that the above thermodynamic metric reproduces the phase transition structure of all four-dimensional
black holes \cite{aqs08}, all known higher dimensional  black holes with and without cosmological constant \cite{qs08},
and generalized three dimensional black holes \cite{qs09}. The main purpose of the present work  is to 
show that the above thermodynamic metric can be used to reproduce correctly the thermodynamics of two dimensional dilatonic 
black holes. This case has been analyzed previously in \cite{twobh} by using a different approach in which Legendre invariance
is not taken into account.

\section{Dilatonic black holes in two dimensions} 
\label{sec:dil}
The two-dimensional Einstein-Hilbert action is just the  Gauss-Bonnet topological term and, therefore, the 
corresponding gravitational models are locally trivial, unless additional matter fields are introduced. 
The most popular models are the generalized dilaton theories which are described by the action (for a recent
review see \cite{gkv02})
\be
I = \frac{1}{8\pi^2}\int d^2 x \sqrt{-h} \left[ X R + U(X) (\nabla X)^ 2 - \lambda V(X)\right] \ ,
\label{actdil}
\ee
where $h = {\rm det}(h_{\mu\nu})$, $R$ is the Ricci scalar corresponding to the metric $h_{\mu\nu}$, 
$X$ is the dimensionless dilatonic field, $U$ and $V$ are arbitrary functions which define the theory, and $\lambda$
is a constant parameter. It can be shown that the general solution to the corresponding field equations 
leads in the Eddington-Finkelstein gauge to the line element
\be
\label{lele}
ds^2 = e^{ Q(X)} \left[ (M + \lambda \omega ) du^2 + 2 du d X \right ] \ , 
\ee
where $ Q'(X) = - U(X),\ \omega'(X) = e^{Q(X)} V(X)$, 
the prime represents differentiation with respect to $X$, and $M$ is a constant of motion that can be interpreted as 
the mass. Clearly, the line element (\ref{lele}) possesses a Killing vector $\xi = \partial_u$ with norm $|\xi| 
= e^{ Q(X)} (M+\lambda \omega)$. Consequently, the solutions of  $M + \lambda \omega(X) =0$ represent Killing horizons
which determine two-dimensional black hole configurations. In this case, it can be shown \cite{gkl95} that the 
Bekenstein-Hawking entropy is 
given by $S=X_h$, where $X_h$ is the ``radius" of the outermost Killing horizon, i.e. $X_h$ satisfies 
the equation $M + \lambda \omega(X_h) = 0$. Furthermore, the  potential $V(X)$ depends on an additional constant parameter
$q$ which is usually interpreted as the dilatonic charge. Then, on the outermost horizon we have that
\be
\label{feq}
M = -\lambda \omega(S, q) \ .
\ee
This equation relates the total mass of the black hole with its entropy and dilatonic charge. If 
we furthermore interpret $M$ as representing the internal energy of the black hole configuration,
then Eq.(\ref{feq}) represents a fundamental equation $M=M(E^a)$ with $E^1= S$ and $E^2=q$
being the extensive variables. As mentioned above, in GTD the fundamental equation
contains all the information that is required to construct the metric of the equilibrium space. Indeed, from Eq.(\ref{gdown})
we obtain the thermodynamic metric
\be
\label{gdowndil}
g = \left(S\frac{\partial M}{\partial S} + q\frac{\partial M}{\partial q}\right)
\left( -\frac{\partial^2 M}{\partial S^2}d S^2 +  \frac{\partial^2 M}{\partial q^2}d q^2\right) \ .
\ee
On the other hand, the condition $\varphi^*(\Theta) = 0$ for $\Theta = dM - T dS - \psi dq $ 
generates the first law of thermodynamics and the equilibrium conditions
\be
T = \frac{\partial M}{\partial S} = \left|\lambda \frac{\partial \omega}{\partial S}\right| = |\lambda \omega'(X)|_{X=X_h} \ , \quad
\psi =   \frac{\partial M}{\partial q} = - \lambda \frac{\partial \omega}{\partial q} \ , 
\ee
where $T$ is the temperature and $\psi$ is the dilatonic potential. This expression for the temperature coincides 
with the result derived from the definition in terms of the surface gravity \cite{kv99}. A useful thermodynamic 
variable is the heat capacity at constant dilatonic charge
\be
C = T \left(\frac{\partial S}{\partial T}\right)_q = \frac{\partial M}{\partial S}\left(\frac{\partial^2 M}{\partial S^2}\right)^{-1}
\ee
the divergencies of which are interpreted in standard black hole thermodynamics \cite{davies} 
as indicating the existence of second--order phase 
transitions. Notice that the determinant of the thermodynamic metric 
\be
{\rm det}(g) = - \frac{\partial^2 M}{\partial S^2} \frac{\partial^2 M}{\partial q^2}
\left(S\frac{\partial M}{\partial S} + q\frac{\partial M}{\partial q}\right)^2 
\ee
vanishes at the points ${\partial^2 M}/{\partial S^2}=0$ where the heat capacity diverges, indicating a possible relationship
between points of phase transitions and zero-volume singularities. In fact, we will show below in concrete examples that phase 
transitions correspond to singularities at the level of the thermodynamic curvature. 

Recently in \cite{kj06}, an interesting symmetry of the action (\ref{actdil}) was found that interchanges the role of the integration 
constant $M$ and the action parameter $\lambda$. Consider the dual variables and functions 
$d\tilde X = d X/\omega(X),\ e^{\tilde Q (\tilde X )} d\tilde X = e^{Q(X)}d X, \ \tilde \omega (\tilde X ) = 1/\omega(X)$ and 
construct the dual action 
\be
\tilde I = \frac{1}{8\pi^2}\int d^2 x \sqrt{-h} \left[ \tilde X R + \tilde U(\tilde X ) (\nabla \tilde X)^ 2 - M \tilde V(\tilde X)\right] \ ,
\label{actdual}
\ee
where $\tilde U (\tilde X) = \omega(X)U(X)- e^{Q(X)}V(X)$ and $ \tilde V (\tilde X) = - V(X)/\omega^2(X)$. Then, by analogy to (\ref{lele}),
the general solution of the field equations following from the dual action (\ref{actdual}) leads to the line element
\be
ds^2 = e^{ \tilde Q(\tilde X)} \left[ (\lambda + M\tilde \omega ) du^2 + 2 du d \tilde X \right ] \ . 
\label{leldual}
\ee    
Since the constant $\lambda$ appears now as an integration constant and $M$ is inside the dual action, the roles of $\lambda$ and $M$ are 
interchanged. Although the dilaton field changes under a dual transformation, the line element remains invariant and represents 
the general solutions to the equations of two different actions. The dual line element (\ref{leldual}) 
has a Killing horizon at $ \lambda + M \tilde \omega(\tilde X) =0$. 
From here it follows the fundamental equation 
\be
\label{dfe}
M=-\frac{\lambda}{\tilde \omega (\tilde S , q)} \ , 
\ee
where $\tilde S = \tilde X _h$ is the value of the dual dilaton field on the outermost horizon.  
According to the description of GTD presented 
in Sec. \ref{sec:int}, the fundamental equation  (\ref{dfe}) completely determines the geometry of the corresponding equilibrium space. 
Then, to determine if a dual transformation leaves invariant the structure of the equilibrium space, it is sufficient to demand 
equivalence between the dual fundamental equation (\ref{dfe}) and the original one given in Eq.(\ref{feq}). 
This condition is satisfied if $\tilde \omega (\tilde S, q) = 1/\omega(S,q)$, i.e., $\tilde\omega (\tilde X_h) = 1/\omega(X_h)$
must hold on the outermost horizons. Since the zeros of $ \lambda + M \tilde \omega(\tilde X) =0$ are also zeros of $M+\lambda\omega(X)=0$, 
a horizon of the original black hole corresponds to a horizon of the dual configuration. In particular, on the original outermost 
horizon we have that $M+\lambda\omega(X_h)=0$, i.e., $\omega(X_h)=-M/\lambda$, whereas on the ``dual" outermost horizon it holds
that  $ \lambda + M \tilde \omega(\tilde X_h) =0$,  i.e., $\tilde \omega(\tilde X_h) =-\lambda/M$. It then follows that 
$\tilde\omega (\tilde X_h) = 1/\omega(X_h)$; this result  is in agreement with the definition of a dual transformation
which demands that $\tilde \omega (\tilde X) = 1/\omega(X)$ in general. Consequently, we have shown 
that the geometric description of dilatonic black hole thermodynamics, using the formalism of GTD, is duality invariant.

We now consider explicit examples of dilatonic black hole configurations which follow from the action (\ref{actdil}). 
For simplicity we choose $\lambda = 2$ and and rescale the mass as $M \rightarrow 2M$. An entire class of black hole 
configurations can be obtained by choosing different values for the potentials $U(X)$ and  $V(X)$ 
(see \cite{gm06} for a review of the most important cases). 
Each choice leads to a specific value for the auxiliary function $\omega(X)$. A family of such solutions
\cite{twobh} is characterized by the fundamental equation ($b\neq -1 \neq c)$
\be
M = \frac{A}{b+1} S^{b+1} + \frac{B}{2(c+1)} S^{c+1} q^2 \ , 
\ee 
which includes the $ab-$family \cite{kkl97} and its Reissner-Nordstr\"om generalization for different 
values of the constants $A$, $B$, $b$, and $c$. The corresponding thermodynamic metric (\ref{gdowndil}) becomes
\be
g= \left[ A S^b + \frac{B(c+3)}{2(c+1)} S^c q^2\right] \left[ -\left(AbS^b + \frac{Bc}{2}S^c q^2\right) dS^2 
+ \frac{B}{c+1} S^{c+2} dq^2\right] \ ,
\label{tmet}
\ee
and the relevant thermodynamic variables are
\be
\label{tvar}
T = \frac{1}{2}(2 AS^b + B S^c q^2) \ , \quad C = \frac{2 A S^{b+1} + B S^{c+1} q^2}{  2 AbS^b + Bc S^c q^2}
= \frac{2ST}{2bT -(b-c)BS^cq^2} \ .
\ee
From the expression for the heat capacity it follows that phase transitions take place at those
 points where $2bT -(b-c)BS^cq^2=0$. Clearly, this equation has nontrivial solutions. The thermodynamic curvature
 of the metric (\ref{tmet}) can be written as
\be 
\label{tcur}
  R = \frac{8A(c+1)^2S^b [ -2A^2b(c+1){\cal P}_1 S^{2b} + B(c+3) S^c q^2(Ab{\cal P}_2 S^b - B {\cal P}_3 S^c q^2)]}
{[2A(c+1)S^b+B(c+3) S^c q^2]^3  [2AbS^b + BcS^cq^2]^2S^2}
\ee
where the ${\cal P}_1$, ${\cal P}_2$, and ${\cal P}_3$ are constant polynomials 
\be
{\cal P}_1 = (b+c)^2 + 8b\ , \quad  {\cal P}_2 = b^2 - 4bc+2b-5c^2-6c\ ,\quad 
{\cal P}_3 = 3 b^2 + bc^2 -3bc + 2c^2 + c^3 \ .
\ee
The first term in squared brackets in the denominator of $R$ is proportional to the conformal factor of (\ref{tmet}) and can be shown \cite{quev07}
to be  also proportional to the thermodynamic potential $M$, by virtue of Euler's identity. Consequently, this term cannot vanish.
The second term in squared brackets coincides with the denominator of the heat capacity (\ref{tvar}). When this term vanishes, 
it can be shown that the numerator of $R$ remains finite.  This means that the divergencies of
$C$ correspond to singularities of $R$. Consequently, the thermodynamic curvature (\ref{tcur}) 
describes the phase transition structure of this particular 
family of dilatonic black holes. 

It is interesting to find out if the thermodynamic metric can be flat. From the above expression for $R$, one can 
see that the choice $b=0$ and  $c=-3$ leads to a vanishing $R$. 
This special case corresponds to the Rindler ground state solution \cite{fr96} for which 
the curvature of spacetime vanishes. 
In a flat  spacetime it is reasonable 
to expect that no thermodynamic interaction exists. The above result reproduces this behavior since thermodynamic curvature is considered in GTD as
a measure of thermodynamic interaction; in the absence of thermodynamic interaction, the thermodynamic curvature should vanish. An additional
solution with flat thermodynamic curvature is obtained for $c=-3$ and ${\cal P}_1=0$. However, it is easy to verify that no real value of $b$ 
satisfies this condition. The only remaining possibility is ${\cal P}_1 = {\cal P}_2 = {\cal P}_3 =0$. Introducing the solution 
for $b^2$ obtained from ${\cal P}_1=0$ into the equation ${\cal P}_2=0$, we obtain $(b+c)(c+1)=0$. Since $c\neq -1$, we have that $b=-c$, a 
solution that, when substituted back in ${\cal P}_1=0$,  
implies that $b=0$. Hence this case corresponds also to the Rindler solution. 
This analysis shows that, except for the case $b = 0$, the above family of dilatonic 
black holes is characterized by a nonvanishing thermodynamic curvature, indicating the presence 
of thermodynamic interaction. 

In two-dimensional gravity additional models are known which correspond to different choices of the potentials $U(X)$ and $V(X)$. 
In each case, the resulting function $\omega(X)$ characterizes the model. We investigated the particular models \cite{gm06} arising from string theory
\cite{string}, Kaluza-Klein reduced gravitational Chern-Simons term \cite{kk1,kk2}, and Liouville gravity \cite{liou} which correspond to 
\be
\omega_{ST} = -2 b^2 X + \frac{b^2 q^2}{8\pi} \ln X \ , \quad
\omega_{CS} = -\frac{1}{8} (q-X^2)^2\ , \quad
\omega_{LG} = \frac{b}{q} e^{qX} \ .
\ee
GTD delivers a particular thermodynamic metric for each case and in all of them we could corroborate that the thermodynamic curvature is nonzero
and its singularities reproduce the phase transition structure which follows from the divergencies of the heat capacity.

\section{Conclusions}
\label{sec:con}

In this work we applied the formalism of GTD to study the thermodynamics of 
dilaton black holes in two dimensions. We considered a family of solutions 
which contains the most representative examples of two-dimensional 
black hole configurations, and found that the (flat) 
Rindler ground state solution is the only solution for which the thermodynamic
curvature vanishes. In all the remaining cases, the singularities of the 
thermodynamic curvature correspond to points where the heat capacity diverges and  phase transitions take place.
We interpret this result as an additional indication that the thermodynamic curvature, as defined in GTD, 
can be used as  measure of thermodynamic interaction. In fact, it has been shown \cite{qsv08} 
that in the case of more realistic thermodynamic systems \cite{callen}, the ideal gas is also 
characterized by a vanishing thermodynamic curvature, whereas the van der Waals gas generates
a nonvanishing curvature whose singularities reproduces the respective phase transition structure.

We analyzed the duality symmetry of dilaton gravity and found the condition for which 
the results of GTD remains invariant under a dual transformation. Furthermore, it was shown that this condition
is always satisfied; we can therefore conclude that GTD is, in general, duality invariant.

\section*{Acknowledgements} 
This work was supported in part by Conacyt, Mexico, grant 48601.  We thank an anonymous referee for 
valuable suggestions.


\end{document}